# THE EARLY UNIVERSE AS A QUANTUM GROWING NETWORK

P. A. Zizzi

Dipartimento di Astronomia dell' Università di Padova
Vicolo dell' Osservatorio, 2
35122 Padova, Italy
zizzi@pd.astro.it

**Abstract**

We consider a quantum gravity register that is a particular quantum memory register which grows with time, and whose qubits are pixels of area of quantum de Sitter horizons. At each time step, the vacuum state of this quantum register grows because of the uncertainty in quantum information induced by the vacuum quantum fluctuations. The resulting virtual states, (responsible for the speed up of growth, i.e., inflation), are operated on by quantum logic gates and transformed into qubits. The model of quantum growing network (QGN) described here is exactly solvable, and (apart from its cosmological implications), can be regarded as the first attempt toward a future model for the quantum World-Wide Web. We also show that the bound on the speed of computation, the bound on clock precision, and the holographic bound, are saturated by the QGN.

## 1. Introduction

It seems to us that fundamental research in theoretical physics has reached a stage where it is impossible to deal with different issues separately: it is time for interdisciplinarity.

General Relativity and Quantum Mechanics have to be entangled in what is called Quantum Gravity (either in the context of String Theories (and M-Theory) [1] or Loop Quantum Gravity [2] and spin networks [3], or, even in a new theory). In this challenging project, also Quantum Cosmology [4], Inflationary Theories [5], Quantum Computation and Quantum Information [6] have to be involved.

Unfortunately, both loop quantum gravity and string theory, do not take into account quantum computing, although it is a peculiar feature of quantum physics in general. David Deutsch says that "quantum computing *is* quantum mechanics", and we fully agree with him. In our opinion, quantum computing should then play a relevant role in a quantum theory of gravitation. One could go even further, and argue that quantum space-time at the Planck scale *is* quantum information' s processing. Space will be then identified with qubits (units of quantum information), and time will be associated with quantum logic gates.

In this paper, in fact, we will describe the quantum inflationary universe as a quantum growing network (QGN) where the nodes (quantum logic gates) are associated with time steps, and the links (qubits) are associated with pixels of area. (We recall that by the Holographic Principle [7], the information about a region of space, is encoded in the bundary surface: one bit for each pixel of area). But this is not the end of the story: as we will see, other issues come in to play, as scale-free models of growing networks [8] (like the World-Wide Web), and cybernetics [9].

The relations among all these issues are rather intricate. For example, the future WWW culd be a quantum growing network, similar to the one described in this paper.

Also, cybernetics should be taken into account, for several reasons: in particular, the "beginning" of the universe can be described in the context of a self-organizing system, where the quantum growing network plays the role of an attractor.

In Section 2, we discuss the relation between the quantized cosmological constant (in a quantum de Sitter space-time [10] describing a inflationary early universe) and quantum information. Quantum fluctuations of the vacuum, then lead to uncertainty in quantum information.

In Section 3, we show that quantum fluctuations of the vacuum correspond to virtual states in the QGN. Those virtual states are operated on by quantum logic gates and transformed into qubits (available quantum information). The presence of virtual states is responsible for the speed up of growth (inflation).

In Section 4, we study the connectivity distribution of our growing quantum network, and find that, although it is exactly solvable, it is not scale-free. Nevertheless, we believe that a version of our model, supplied with entangled qubits, could furnish a quantum scale-free growing network describing the future quantum WWW.

In Section 5, we show that our quantum growing network satisfies the bounds on the speed of computation and clock precision [11] and also the holographic bound [7].

Section 6, is devoted to some concluding remarks.

## 2. The early universe and quantum information

By assuming that time is quantized [10]:

(1) $t_n = (n+1)t_P$, (with n=0,1,2,…),

where $t_P = \left(\frac{\hbar G}{c^5}\right)^{1/2} \cong 10^{-44}$ sec is the Planck time, it is possible to avoid the initial singularity of the Big Bang. In fact at the initial time $t = t_0 = t_P$, the area of the cosmological horizon, $A_0$, is different from zero, and proportional to $l_p^2$, where $l_p$ is the Planck length: $l_P = ct_P = \left(\frac{\hbar G}{c^3}\right) \cong 1.6\times 10^{-33} cm$.

In [10], a time slicing of a 4-dimensional Riemmannian space-time, was performed, by the use of the quantized time $t_n$ in eq. (1). This leads to a discrete spectrum of the quantum fluctuations of the metric:

(2) $(\Delta g)_n = \frac{1}{n+1}$

and to a discrete spectrum of the area of each spatial slice:

(3) $A_n \cong (\Delta x)^2 g_n = l_n^2$

where:

(4) $g_n \equiv g_{ij}(\vec{x}, t_n)$ (with i,j=1,2,3)

are the spatial components of the metric at time $t_n$, and

(5) $l_n = ct_n = (n+1)l_P$

is the proper length at time $t_n$. The resulting cosmological model is a quantum de Sitter universe, whose cosmological horizon has the discrete area spectrum:

(6) $A_n = 4\pi(n+1)^2 l_P^2$.

A similar result for the area spectrum of quantum de Sitter horizons was found by Schiffer [12], although he used a different approach.

In [10], avoiding the Big Bang singularity was due mainly to the quantization of time. However, a diffent singularity can be avoided without requiring quantization of time: the zero black hole area at the end of evaporation. This can be done by assuming that the area of a quantum black hole has a discrete spectrum [13]: $A_n = \alpha n l_P^2$ (n=1,2,3) where $\alpha$ is a real constant.

In our picture, the quantum fluctuation of the metric $(\Delta g)_n$, turns out to be closely related to the quantum information $I$, stored in the spatial slice, or better, due to the holographic principle [7], in the surface area of the cosmological horizon bounding it. In fact, the following relation holds:

(7) $(\Delta g)_n^{-2} = I$.

Morover, the following relation holds between the quantized cosmological constant [10] and quantum information:

(8) $\Lambda_n = \frac{1}{I\, l_P^2}$.

The quantum information $I$ is the number N of qubits in a quantum memory register , which, in our case, as it was shown in [14], is equal to the number of pixels (units of Planck area) of the quantum de Sitter horizon:

(9) $I = N \equiv (n+1)^2$.

In [10], the positive cosmological constant was quantized, but its relation with quantum information became clear only after having considered the aspects of the quantum holographic principle in [14]. Indipendently, Banks [15] claimed that the cosmological constant determines the number of degrees of freedom in an asymptotically de Sitter universe.

Also, very recently, Bousso [16] argued that the total observable entropy is bounded by the inverse of the cosmological constant, and that this fact holds for all space-times with a positive cosmological constant. He calls this: "The N-bound", where N is the number of degrees of freedom.

The quantum entropy of the $n^{th}$ de Sitter horizon is:

(10) $S_n = N \ln 2$.

At the initial time (the Planck time), the quantum fluctuation of the metric and the cosmological constant get their maximum values:

(11) $(\Delta g)_0 = 1; \; \Lambda_0 = \dfrac{1}{l_P^2}$,

while quantum information, and quantum entropy, get their minimum values:

(12) $I = 1; \; S_0 = \ln 2$.

The quantum de Sitter horizon, at the Planck time, coincides with the horizon of a Planckian black hole, whose area is one pixel:

(13) $A_0 = 4\pi l_P^2$,

encoding 1 qubit of information..

From eq. (8), it follows that a quantum fluctuation of the vacuum, i.e, uncertainty in the vacuum energy $\Delta\Lambda$ corresponds to uncertainty in the quantum information $\Delta I$.

In fact, we have:

(14) $\Delta I = \dfrac{-\Delta\Lambda}{l_P^2 \Lambda^2} = 2n+3$

The 2n+3 vacuum states in eq. (14) will be interpreted as virtual states in the QGN.

From eq.14, if $\Delta\Lambda < 0$, as in the case: $\Delta\Lambda = \Lambda_{n+1} - \Lambda_n < 0$, we have:

(15) $\Delta I_{n'n} = I_{n+1} - I_n = 2n+3 = 2n'+1$ (where n'=n+1)

that is, an increase of quantum information. In fact, from the point of view of an observer on the horizon $n'$ (with respect to the horizon n), there is an increase of information, coming from the preceeding horizon n.

If instead $\Delta\Lambda > 0$, as in the case: $\Delta\Lambda = \Lambda_n - \Lambda_{n+1} > 0$, we have:

(16) $\Delta I_{nn'} \equiv I_n - I_{n+1} = -(2n+3)$

Then, from the point of view of an observer on the horizon n with respect to the horizon $n' = n+1$, the virtual states are black holes where information is lost.

The quantum entropy $S_n$ of the horizon n, which is the quantum entropy of N qubits, with $N \equiv (n+1)^2$, namely: $S_n = (n+1)^2 \ln 2$, is the sum of the increases of entropy , from all preceeding horizons:

(17) $S_n = \Delta S_{-1} + \Delta S_0 + \Delta S_1 + \Delta S_2 + ... + \Delta S_{n-1}$,

where:

(18) $\Delta S_n = \Delta I_{n'n} \ln 2$ .

For example, for n=2, we have: $S_1 = 4\ln 2 = \Delta S_{-1} + \Delta S_0$ where $\Delta S_{-1} = \ln 2$ and $\Delta S_0 = 3\ln 2$ .

## 3. Virtual states in the quantum gravity register

A quantum memory register of size n is a collection of n qubits. Information is stored in the quantum register in binary form.

The state of n qubits is the unit vector in the $2^n$-dimensional complex Hilbert space:

$C^2 \otimes C^2 \otimes ... \otimes C^2$ n times.

As a natural basis, we take the computational basis, consisting of $2^n$ vectors, which correspond to $2^n$ classical strings of length n:

$|0\rangle \otimes |0\rangle \otimes ... \otimes |0\rangle \equiv |00...0\rangle$

$|0\rangle \otimes |0\rangle \otimes ... \otimes |1\rangle \equiv |00...1\rangle$

.

.

.

$|1\rangle \otimes |1\rangle \otimes ... \otimes |1\rangle \equiv |11...1\rangle$

In general, we will denote one basis vector of the state of n qubits as:

$|i_1\rangle \otimes |i_2\rangle \otimes ... \otimes |i_n\rangle \equiv |i_1 i_2 ... i_n\rangle \equiv |i\rangle$

where $i_1, i_2, ..., i_n$ is the binary representation of the integer i, a number between 0 and $2^{n-1}$. For example, a quantum register of size 4 can store numbers such as 13 or 9:

$|1101\rangle = |13\rangle$

$|1001\rangle = |9\rangle$

The general state is a complex unit vector in the Hilbert space, which is a linear superposition of the basis states:

$$\sum_{i=0}^{2^n-1} c_i |i\rangle$$

where $c_i$ are the complex amplitudes of the basis states $|i\rangle$, with the condition:

$$\sum_i |c_i|^2 = 1$$

To perform computation with qubits, we have to use quantum logic gates.

A quantum logic gate on n qubits is a $2^n \times 2^n$ unitary matrix U.

The unitary matrix U is the time evolution operator which allows to compute the function f from n qubits to n qubits:

$|i_1 i_2 ... i_n\rangle \rightarrow U|i_1 i_2 ... i_n\rangle = |f(i_1 i_2 ... i_n)\rangle$ .

An ensemble of n quantum logic gates is called a quantum network of size n.

Let us consider our Hilbert space, which has dimension $2^{(n+1)^2}$ at time $t_n = (n+1)t_P$ .

At time $t_0 = t_P$, the computational basis consists of 2 vectors (2 classical strings of length 1):

$|0\rangle , |1\rangle$

At time $t_1 = 2t_P$, the computational basis consists of 16 strings of length 4:

$|0000\rangle, |0001\rangle \dots |1111\rangle$

At time $t_2 = 3t_P$ , the computational basis consists of 512 strings of length 9:

$|000000000\rangle$, $|000000001\rangle \dots |111111111\rangle$

and so on. In general, at time $t_n = (n+1)t_P$, the computational basis consists of $2^{(n+1)^2}$ strings of length $(n+1)^2$ .

Thus, at each time step, there is an increase of 2n+3 qubits in the quantum gravity register.

This can be seen from the expression of the discrete evolution operator [14]: $E_{nm} = |\bar{1}\rangle^{(m-n)(m+n+2)}$ in the case m=n+1:

(19) $E_{n,n+1} = |\bar{1}\rangle^{2n+3}$.

where the state $|\bar{1}\rangle$ in eq. (19) is the one-qubit in the equal superposition of the basis states $|0\rangle$ and $|1\rangle$:

(20) $|\bar{1}\rangle = \frac{1}{\sqrt{2}}(|0\rangle + |1\rangle)$

and should not be confused with the basis state $|1\rangle$ .

In the following, we will denote with $|\bar{4}\rangle$, $|\bar{9}\rangle$, $|1\bar{6}\rangle$, $|2\bar{5}\rangle$,... $|\bar{N}\rangle$ the tensor products $|\bar{1}\rangle^{\otimes 4}$, $|\bar{1}\rangle^{\otimes 9}$, $|\bar{1}\rangle^{\otimes 16}$, $|\bar{1}\rangle^{\otimes 25}$, $\dots|\bar{1}\rangle^{\otimes N}$ respectively. These states are called product states or separable states (in the sense that they are not entangled).

In particular, in this case, they are symmetric states.

The bar on the top of the figure is used to distinguish the number $|N\rangle$ in binary representation from the product state $|\bar{N}\rangle = |\bar{1}\rangle^{\otimes N}$, where we recall that $N = (n+1)^2$ .

It should be reminded that the one-qubit state $|\bar{1}\rangle$ can be obtained by the action of the Hadamard gate (Had) on the basis state $|0\rangle$:

$Had|0\rangle = |\bar{1}\rangle$, where: $Had = \frac{1}{\sqrt{2}}\begin{pmatrix} 1 & 1 \\ 1 & -1 \end{pmatrix}$.

The initial state or "imput" of a quantum register is generally taken to be a n string where all qubits are "cooled" in the basis state $|0\rangle$ . This is called the vacuum state of the quantum register:

$|0000000...0\rangle$

In our case, the vacuum state grows at each time step, by an amount of 2n+3 states $|0\rangle$, because of eq. (14).

Let Had(j) represent the Hadamard gate acting on bit j, and :

(21) $U = \prod_{j=1}^{N=(n+1)^2} Had(j)$ .

Let us indicate with $v_n$ the number of virtual states at time $t_n$:

(22) $v_n \equiv 2n+3$

Also, we shall indicate with $|virt\rangle_n$ and $|vac\rangle_n$ the virtual states and the vacuum states respectively. Also, let us define:

(23) $U_n = \prod_{j=1}^{v=2n+3} Had(j)$ with n=0,1,2…

The N qubits at time $t_n$ are given by the application of the operator U in eq. (21) to the vacuum:

(24) $|\overline{N}\rangle = U|vac\rangle_n$

At each time step $t_n$, the virtual state $|virt\rangle_{n-1}$ occurring at time $t_{n-1}$ is transformed into v=2n+3 qubits by the operator $U_n$ in eq. (23):

(25) $U_n|virt\rangle_{n-1} = |\overline{1}\rangle^{\otimes v}$.

The operators $U_n$ will be interpreted as the nodes "n" of the growing quantum network.

At the "unphysical" time $t_{-1}$ ( N=0), it is, by definition: $|vac\rangle_{-1} = 1$, and $U_{-1} = 1$ (the node "-1" is the only unactive node). From eq.(22) we get $v_{-1} \equiv 1$, then the virtual state is: $|virt\rangle_{-1} = |0\rangle$.

At time $t_0 = t_P$, (N=1), we have: $|vac\rangle_0 = |0\rangle$. That means that the virtual state at time $t_{-1}$, turned into a vacuum state at time $t_0$:

$|vac\rangle_0 = |virt\rangle_{-1} = |0\rangle$.

At node "0", the virtual state $|virt\rangle_{-1} = |0\rangle$ (which in this case coincides with the vacuum state $|vac\rangle_0 = |0\rangle$) is operated on by the operator $U_0$ and transformed into one qubit:

$U_0|0\rangle = Had|0\rangle = |\overline{1}\rangle$.

From eq.(22), we get: $v_0 \equiv 3$, then we have: $|virt\rangle_0 = |000\rangle$.

At time $t_1 = 2t_P$, (N=4), we have: $|vac\rangle_1 = |0000\rangle$. From eq.(22) we get: $v_1 = 5$, then we have: $|virt\rangle_1 = |00000\rangle$.

$|vac\rangle_1$ can be written as: $|vac\rangle_1 = |virt\rangle_0 \otimes |vac\rangle_0 = |virt\rangle_0 \otimes |virt\rangle_{-1}$.

At node "1", the virtual state $|virt\rangle_0$ is operated on by the operator $U_1$ and transformed into 3 qubits:

$U_1|virt\rangle_0 = H(1)H(2)H(3)|000\rangle = |\overline{1}\rangle^{\otimes 3}$.

The 4 qubits at time $t_1$ are given by the application of the operator U to the vacuum state:

$U|vac\rangle_1 = \prod_{j=1}^{4} H(j)|0000\rangle = |\overline{1}\rangle^{\otimes 4} = |\overline{4}\rangle$, which can also be written as:

$$U|vac\rangle_1 \equiv \prod_{j=1}^{3} H(j)|000\rangle \otimes H|0\rangle \equiv U_1|virt\rangle_0 \otimes U_0|virt\rangle_{-1} \equiv |\overline{1}\rangle^{\otimes 3} \otimes |\overline{1}\rangle = |\overline{4}\rangle .$$

At time $t_2 = 3t_P$, (N=9), we have: $|vac\rangle_2 = |000000000\rangle$. Also, it is: $v_2 = 7$, thus we get: $|virt\rangle_2 = |0000000\rangle$.

At node "2" the virtual state $\left|virt\right\rangle_1 = \left|00000\right\rangle$ is operated on by the operator $U_2$ and transformed into 5 qubits:

$U_2\left|virt\right\rangle_1 = H(1)H(2)H(3)H(4)H(5)\left|00000\right\rangle = \left|\bar{1}\right\rangle^{\otimes 5}$.

The 9 qubits at time $t_2$ are given by the application of the operator U to the vacuum state:

$U\left|vac\right\rangle_2 = \prod_{j=1}^{9} H(j)\left|000000000\right\rangle = \left|\bar{1}\right\rangle^{\otimes 9} = \left|\bar{9}\right\rangle$, which can also be written as:

$$U\left|vac\right\rangle_2 \equiv \prod_{j=1}^{5} H(j)\left|00000\right\rangle \otimes \prod_{j=1}^{3} H(j)\left|000\right\rangle \otimes H\left|0\right\rangle \equiv U_2\left|virt\right\rangle_1 \otimes U_1\left|virt\right\rangle_0 \otimes U_0\left|virt\right\rangle_{-1} =$$

$$= \left|\bar{1}\right\rangle^{\otimes 5} \otimes \left|\bar{1}\right\rangle^{\otimes 3} \otimes \left|\bar{1}\right\rangle = \left|\bar{9}\right\rangle$$

In general, the N-qubits state at time $t_n$ can be written as:

(26) $\left|\bar{N}\right\rangle = U_n\left|virt\right\rangle_{n-1} \otimes U_{n-1}\left|virt\right\rangle_{n-2} \otimes ...U_1\left|virt\right\rangle_0 \otimes U_0\left|virt\right\rangle_{-1}$

The quantum algorithm is illustrated by the following family of quantum networks. The diagrams below provide a schematic representation of each quantum network, where H represents the Hadamard gate.

At time $t_0$, we have a quantum network of size one:

(27) $\left|0\right\rangle$ ____ $H$ ______ $\dfrac{\left|0\right\rangle + \left|1\right\rangle}{\sqrt{2}}$

At time $t_1$, we have a quantum network of size four:

(28)
$$\left.\begin{matrix} \left|0\right\rangle \text{____ } H \text{ ______ } \dfrac{\left|0\right\rangle + \left|1\right\rangle}{\sqrt{2}} \\ \left|0\right\rangle \text{____ } H \text{ ______ } \dfrac{\left|0\right\rangle + \left|1\right\rangle}{\sqrt{2}} \\ \left|0\right\rangle \text{____ } H \text{ ______ } \dfrac{\left|0\right\rangle + \left|1\right\rangle}{\sqrt{2}} \\ \left|0\right\rangle \text{____ } H \text{ ______ } \dfrac{\left|0\right\rangle + \left|1\right\rangle}{\sqrt{2}} \end{matrix}\right\} = \frac{1}{4}(\left|0000\right\rangle + \left|0001\right\rangle + \left|0010\right\rangle + ...\left|1111\right\rangle)$$

And so on. In general, at time $t_n$, the quantum network has size $N = (n+1)^2$.

## 4. The quantum growing network

Random networks, growing networks, and quantum networks, are three different kind of networks.

In a random network, the number of nodes is kept fixed, and the links among nodes are distributed randomly.

In a growing network, the number of nodes grows with time, and in the particular case of scale-free growing networks, there is preferential attachment, i.e., new nodes attach preferentially to already well connected nodes. This is the case of the World-Wide Web, for example. Theoretical models [8] describing such kind of growing networks, show that the connectivity distribution of nodes follows a power-law of the kind: $P(k) \approx k^{-\gamma}$ where k is the connectivity, and the exponent $\gamma > 2$. In particular, the experimental result for the WWW, is $\gamma = 2.1 \pm 0.1$.
Finally, quantum networks, are networks of quantum logic gates, but in general, they are not growing with time.
However, the quantum network discussed in section 3, is a growing network, and in what follow, we will look for its connectivity distribution in order to find which kind of growing network it is (scale-free or not).
The rules of the growing quantum network that we consider, are resumed below.
At the starting time (the unphysical time $t_{-1} = 0$), there is one node, call it **-1**. At each time step $t_n$, a new node is added, which links to the youngest and the oldest nodes, and also carries 2n+1 free links. Thus, at the Planck time $t_0 = t_P$, the new node **0** is added, which links to node **-1** and carries one free link**.** At time $t_1 = 2t_P$, the new node **1** is added, which links to nodes **-1** and **0,** and carries three free links. At time $t_2 = 3t_P$, the new node **2** is added, which links to nodes **-1** and **1**, and carries five free links. At time $t_3 = 4t_P$, the new node **3** is added, which links to nodes **-1** and **2**, and carries seven free links, and so on. See fig.1.
In general, at time $t_n$, there are:

1) $N^* = n + 2$ nodes, but only n+1 of them are active, in the sense that they have outgoing free links (node **-1** has no outgoing free links).
2) $N = (n+1)^2$ free links coming out from n+1 active nodes
3) 2n+1 links connecting pairs of nodes
4) n loops.

The N free links are qubits (available quantum information), the 2n+1 connecting links are virtual states, carrying information along loops, the n+1 active nodes are quantum logic gates operating on virtual states and transforming them into qubits. In fact, notice that the number of free outgoing links at node **n** is 2n+1, which is also the number of virtual states (connecting links) in the loops from node **-1** to node **n .**
As there is one node per Planck time unit, a Planck time unit can be identified with a Hadamard gate. Thus, quantized time allows the transformation of virtual quantum information into real quantum information.
The connected part of the network in fig.1 is the most similar to a scale-free growing network, where free links are absent. However, the connected part is deterministic, in the sense that it follows some precise rules, as a lattice. What is missing here, is one of the two fundamental features of scale-free networks, which is preferential attachment. The free links, however, destroy the structure of a regular lattice, as the configuration of free links changes at each time step.
In what follows, we will investigate the connectivity distribution of this growing network.
The connectivity k(**i)** of a node (or site) **i,** is the number of his connections (links connecting the site **i** to other sites). By definition, then, free links coming out from a site **i** cannot be included in the connectivity, although they are included in the degree of the node (the degree of a node is the total number of links at a node).
As in our case the links are directed, we will consider both the connectivity of incoming links, $k_{in}(i)$ and the connectivity of outgoing links, $k_{out}(i)$.
The node **-1** has got n+1 incoming links. So, the in-connectivity of node **-1** is: $k_{in}(-1) = n + 1$. The last node **n** has no incoming links, then it is: $k_{in}(n) = 0$. All other nodes **i** have one incoming link, so their in-connectivity is:

(29) $k_{in}(i) = 1$ (where i=0,1,2…n-1).

The node **-1** has no outgoing links, so his out-connectivity is zero: $k_{out}(-1) = 0$. The node **0** has one outgoing link: $k_{out}(0) = 1$. All other nodes **i** have two outgoing links: $k_{out}(i) = 2$.

The connectivity distribution P(k) gives the probability that a node in the network is connected to k other nodes. When the links are directed, one has to consider both $P(k_{in})$ and $P(k_{out})$.

In our case we have:

(30) $P(k_{in}) \cong \delta(k-1)$

(as in the limit of a large network we can disregard the first and the last nodes), and:

(31) $P(k_{out}) \cong \delta(k-2)$

(by neglecting nodes **-1** and **0** in the thermodynamic limit).

Finally, although it is not possible to speak about a proper connectivity for free links, one could ask which is the probability that a whatever node of the network has $k_n = 2n+1$ outgoing free links (n=0,1,2,3…). One finds that this probability is uniform:

(32) $P(k_{out-free}) = 1/N^*$

where we recall that $N^* = n+2$ is the total number of nodes at time $t_n$.

We notice that none of the connectivity distributions of our network is a power law of the kind $P(k) = k^{-\gamma}$. In other words, our growing quantum network is not scale-free. This is due to the fact that, although this network is a growing one, it lacks the second very important requirement to be scale-free, which is preferential attachment. It might be interesting to look for scale-free quantum networks. The question is whether the future World Wide Web will be indeed a quantum growing network. Most probably, a scale-free quantum network describing the quantum WWW, should use entangled qubits, (differently from the product states used in the groving quantum network discussed in this paper. In fact, very recently, Todd Brun [17] described a quantum web page as a generalization of teleportation to N parties sharing a N-qubit entangled state.

## 5. Saturated bounds

Interestingly, the QGN saturates the bound on the speed of computation $\nu$ [11]:

$$I\nu^2 \leq \frac{1}{t_P^2}$$

In fact, as in our case it is $I = N = (n+1)^2$, we have:

(33) $$\nu_n = \frac{1}{\sqrt{N}t_P} = \frac{1}{(n+1)t_P} = \frac{1}{t_n}$$

Moreover, the QGR saturates the bound on clock precision [11]:

$$T \leq t\left(\frac{t}{t_P}\right)^2$$

where t is the accuracy (the minimal time that can be used), and T is the total running time.

In our case, time is quantized in Planck time units, so that the minimal time that can be used is the Planck time, then we have:

(34) $t = t_P = T$

Finally, let us consider the bound on distance inaccurancy $\delta R$ [11] in the measurement of a distance R between two points:

$$\delta R\left(\frac{\delta R}{l_P}\right)^2 \geq R$$

Since in our case it is: $R_n \approx (\Delta g)_n^{-1} l_P$, and $\delta R_n \approx (\Delta g)_n^{-2} l_P$, it follows that:

(35) $R_0 = \delta R_0 = l_P$

as $(\Delta g)_0 = 1$, and the bound is saturared at the Planck scale, for n=0.

The bound on distance inaccuracy leads to the holographic bound [7] as it was shown in [11]:

$$N \leq \left[\frac{R}{(Rl_P^2)^{1/3}}\right]^3.$$

where N is the number of degrees of freedom of the region of linear dimension R, and $\left[\frac{R}{(Rl_P^2)^{1/3}}\right]^3$ is the area of the region in Planck units.

In our case it is: $R \equiv l_n = (n+1)l_P$, then we get:

(36) $$\left[\frac{R}{(Rl_P^2)^{1/3}}\right]^3 = \left[\frac{(n+1)l_P}{\left((n+1)l_P^3\right)^{1/3}}\right]^3 = (n+1)^2 = N = I .$$

## 6. Concluding remarks

i) Attractors, are local minima of the potential function in the state space of a system, or, which is the same, local maxima of the Fitness function, which is the negative of the potential function. Fitness is a measure of the stability of a system, and/or its potential for growth. Thus, there are attactors with higer or lower fitness.

One could argue that the QGN discussed in this paper, is one of the attractors of some self-organizing system. That self-organizing system might be some kind of non local and non causal space-time structure made up of entangled qubits [18]. Although that system does not represent any physical space-time, it can be considered as a ***proto space-time***, which is the seed of physical quantum space-time. To some extent, we agree with Manfred Requardt [19], who claims that at the Planck scale, space-time is a discrete substratum described by a cellular (random) network encoding non-local aspects of quantum physics. However, we think that such a random and non-local structure exists just below the Planck scale. At the Planck scale, the random network has already self-organized into the QGN. Indeed, we believe that the quantum beginning of physical space-time took place at the node **0** (the Hadamard gate) of the QGN. Quantized time appears as the result of the transformation of virtual states (vacuum energy) into qubits (quantum information) at the nodes of a quantum network. This interpretation of time, is very much in agreement with the idea of Scott Hitchocck [20], that time is the result of the conversion of energy into information.

ii) The speed up of the growth of the QGN, is due to virtual states and it is responsible for quantum inflation. If virtual states were absent in the quantum network, the growth would be much slower. In that case, the early universe could be interpreted as a $2^n$ lattice (n=0,1,2…), represented by the regular tree graph in Fig.2.
iii) The philosophical interpretation of virtual states in the quantum growing network, is rather intriguing: the early universe, during (quantum) inflation, is passing through alternating states of propensity (virtual states) and actuality (qubits).
When decoherence occours [14], at the end of inflation, qubits collapse to classical bits, and virtual states are not present anymore. At this point the universe settles down in a permanent state of actuality ("existence") because of the absence of virtual states.
This seems to be the second attractor where the system settles down, and it must be a fitter one, in fact the system becomes very stable, because classical bits, unlike qubits, obviously do not undergo decoherence. What happens is that noise, in this case thermal noise, pushes the system toward a higer potential (lower fitness), then the system escapes from the previous attractor, and has the chance to end in a fitter attractor.
iv) According to this picture, the Big Bang was not the true beginning of existence, but just a chance for it.
Then, we have three "degrees" (or phases) for the beginning of the universe:
-The (perhaps eternal) presence of the proto space-time below the Planck scale.
-The beginning of (quantum) inflation at the Planck time.
-The end of inflation and beginning of existence.
v) Quite recently, cybernetic research focused on the study of an intelligent Web [21] (where "intelligent" in reality means "learning", and morover, the process of learning in the Web takes place by the aid of the users). The learning process in the Web is very similar to the one taking place in the human brain: a process of associative learning (or Hebbian learning). Thus, a link between two hypertexts (nodes) becomes stronger and stronger, the more frequently it is used, like associations in the brain.
We believe that a future quantum Web will employ faster search engines, using the Grover algorithm [22] (if a classical computer is able to search n items in a certain time t, the Grower algorithm allows a quantum computer to search a total of $n^2$ items in the same time).
vi) A quantum Web could even undergo conscious experiences, if we believe Penrose and Hameroff, who claim that our mind is very similar to a quantum computer, and that conscious experiences are due to decoherence of tubulins-qubits [23].
The idea of a conscious quantum Web is quite in agreement with the Global Brain idea [24 ], which foresees the Web in symbiosis with his users (the Net becoming the brain of a superorganism of which humans are just a component).
vii) The beginning of existence of the universe (at the end of inflation due to decoherence) coincided with a cosmic conscious event [25] of which our brain structure is still reminiscent.

## Aknowledgments

It is a pleasure to thank Ginestra Bianconi for enlightening discussions on some technicalities of scale-free growing networks, for a great deal of e-mail exchange, help, and encouragement. I am very grateful to Robert Tucci for very useful discussions on quantum logic gates. I also wish to thank Albert-Laszlo Barabasi for helpful comments and advice.

**Fig.1: The early universe as a quantum growing network. Virtual states are responsible for the speed up of growth (inflation).**

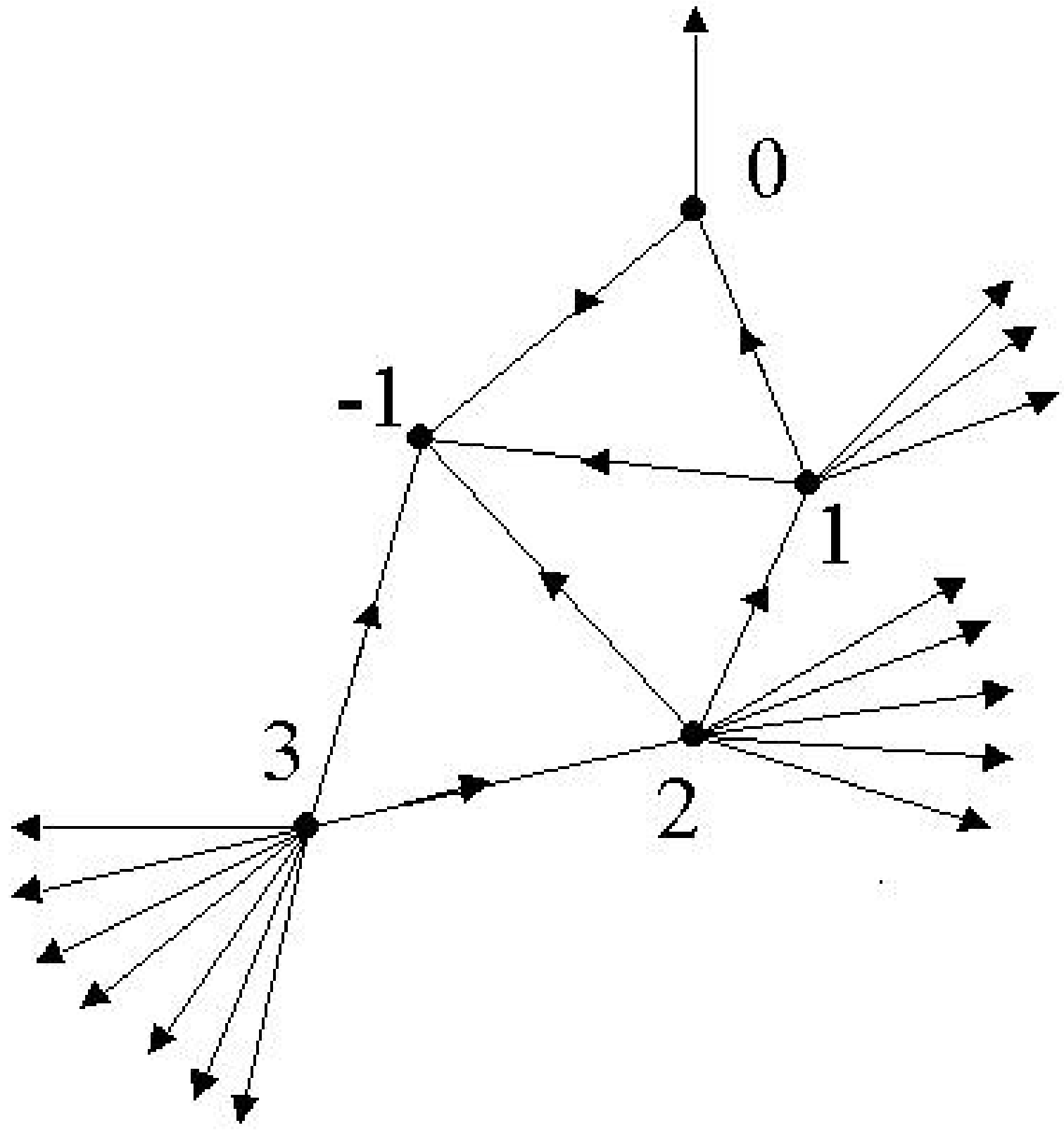

**Fig.2: If virtual states were absent, the early universe would be represented by a regular lattice, but the growth would be much slower.**

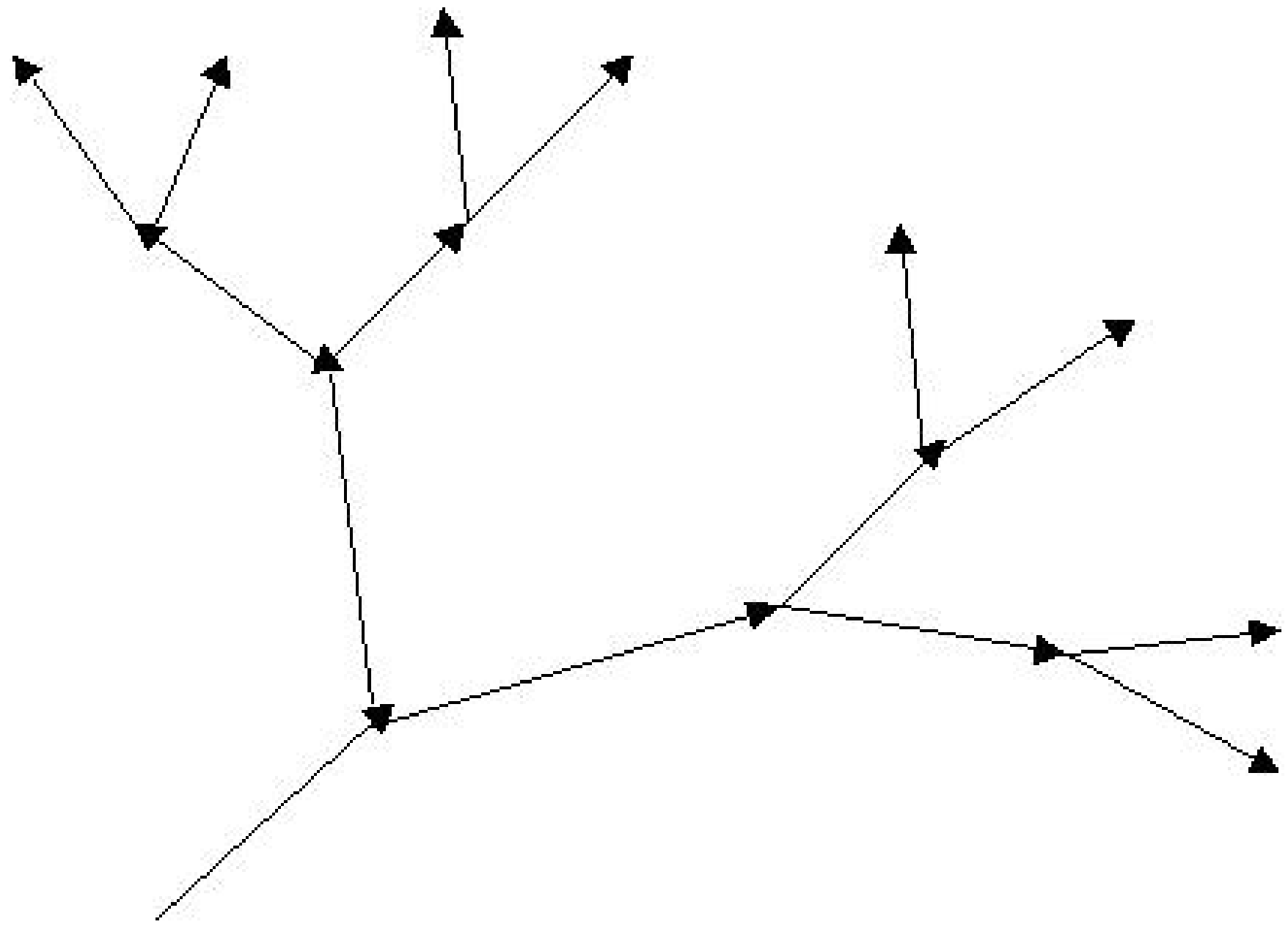